\documentclass[12pt]{article} \textheight 23cm
\usepackage{graphicx}
\usepackage{amsmath}
\usepackage{color}

\begin{document} \begin{center}
{\large \bf Multi-scale Patterns formed by Sodium Sulphate in a Drying Droplet of Gelatin
} \vskip 0.2cm
Biswajit Roy$^1$, Moutushi Dutta Choudhuri$^1$, Tapati Dutta$^2$ and Sujata Tarafdar$^{1*}$\\
\vskip 0.2cm

$^1$ Condensed Matter Physics Research Centre, Physics Department, Jadavpur University, Kolkata 700032, India\\

$^2$ Physics Department, St. Xavier's College, Kolkata 700016, India\\

Sujata Tarafdar$^*$: Corresponding author, sujata$\_$tarafdar@hotmail.com, Phone 913324146666(Ex. 2760), Fax: 913324138917 

\end{center}
\vskip .2cm

 \noindent	Abstract\\ 
 \noindent
 We present a study of patterns, formed in drying drops of aqueous gelatin solution containing sodium sulphate. The patterns are highly complex, consisting of a hierarchical sequence of rings which form concentric bands as well as  dendritic crystalline aggregates. When the preparation of the complex fluid is done by mixing an aqueous solution of the salt with an aqueous solution of gelatin prepared separately, another feature is observed in the  pattern on the dried out drop. This is a viscous fingering pattern, superposed on the series of rings. We try to explain the origin of these two features from a simple physical approach.
 
 \section{Introduction}
 The study of crystallization patterns of salts in drying droplets shows a wide variety of intricate and beautiful patterns$^{1,2}$. Sodium sulphate and sodium chloride, in high viscosity solutions are popular candidates for such studies. Crystal formation on macroscopic scales to microscopic dendritic growth have been reported$^{3}$, ring formations like the well-known coffee stain$^{4}$ and multiple rings have also been observed$^{5}$. Many different aspects of drying droplets are currently under study$^{6,7}$.
 
 Sodium sulphate is a constituent of Portland cement and usually crystallizes as  thenardite (Na$_2$SO$_4$) or mirabilite (Na$_2$SO$_4$.10H$_2$O)$^{8}$ depending on ambient conditions and water content. Several different polymorphic forms have been observed.
Rodriguez-Navarro et al.$^{8}$ further show that crystallization pressure developed within porous materials when thenardite is formed, can cause more damage to building material and rocks than formation of mirabilite. A metastable form, sodium hepta-hydrate (Na$_2$SO$_4$.7H$_2$O) is also believed to be significant in this context$^{8,9}$.
So such studies are important from a practical point of view.
 
  If the salt solution in pure water is dried, a coffee-ring pattern$^4$ is observed. On highly hydrophillic surfaces, the salt forms dendritic patterns \textit{outside} the drop initially deposited$^{8}$. This is due to a precursor wetting film surrounding the drop. The anhydrous  salt crystallizes in rhombohedral form. If the salt is dissolved in a complex viscous fluid e.g. containing a polymer, the dried drop shows more complicated features$^{5}$.

 In the present paper we focus primarily on two interesting observations - (i) a hierarchical series of concentric  rings whose distribution is similar to a devil's staircase$^{10}$ and (ii) a viscous fingering pattern$^{11}$ superposed on the rings, which is observable under certain conditions.
 	
 \section{Materials and Methods}
 The complex solution for droplet deposition was prepared by two slightly different methods and the results showed a noticeable difference in the pattern left on the substrate after the droplet dries.
 \subsection{Method I} 
 0.142 g sodium sulphate powder was mixed with  50 ml. HPLC grade water, to get a solution of 0.02M concentration.
0.5 g gelatin powder was mixed with this solution and stirred at temperature 60$^\circ$C until completely dissolved. 
 
 \subsection{Method II}
 0.2 g of gelatin was added to 20 ml of HPLC grade water, this gives a concentration of 1\% by weight. It was stirred in a magnetic stirrer at temperature of 60$^\circ$C until it completely dissolves.
 
 Sodium sulphate solution in water was prepared by mixing $x$ g of salt in 20 ml. water, where $x$ = 0.028, 0.056, 0.085, 0.113 and 0.142  respectively.
 These two solutions were mixed in 1:1 ratio by volume. The molar concentration of the salt was varied from 0.01 to 0.05 M.
 
 A droplet of each solution of volume $\sim$ 50$\mu$l was deposited using a micro-pipette on a clean glass surface and allowed to dry under ambient conditions.
 
 \section{Results}
 The appearance of the drops on drying and some relevant physical properties were measured.
 Results are given below.
\subsection{Physical properties: Surface tension, Viscosity and angle of contact measurement}
 Surface tensions of (i) sodium sulphate in water, (ii) gelatin in water and (iii) the fluid containing both salt and gelatin, prepared by method II, were measured by KRUSS Tensiometer K9 at Physics Department, Jadavpur University.
Viscosity of the above three solutions was measured by Brookfield DV-III Ultra viscometer at Gurudas College, Kolkata. 
The angles of contact of sample (iii) on glass was measured to be approximately 18 $\pm 2^\circ $.
 All results are given in Table (\ref{prop}) 
 
 \subsection{Imaging}Gross features of the pattern on the dried drop were visible to the naked eye, but new features were revealed  as the resolution of observation changed.
 \subsubsection{Low resolution imaging}
\begin{figure}[h]
\begin{center}
\includegraphics[width=10.0cm, angle=0]{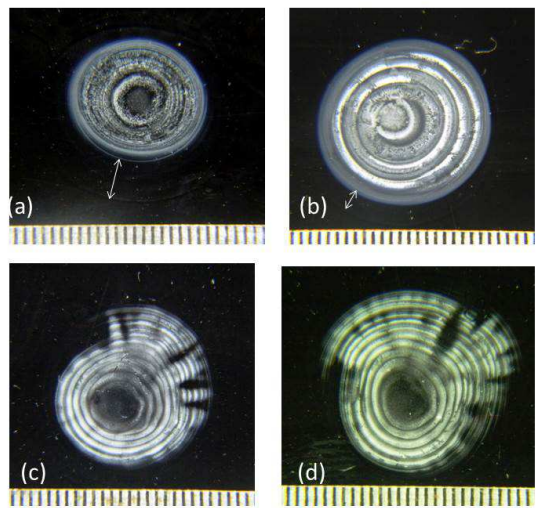}
\end{center}
\caption{The appearance of the dried droplets are shown, the graduations on the scales below are in mm. (a) and (b) are prepared by Method-I, while (c) and (d)  by Method-II  show viscous fingering. (a) is for salt concentration 0.01 M and (b) for 0.03 M. (c) is for 0.02 M solution of 50 $\mu l$ volume, while (d) is for 0.02 M solution of 40 $\mu l$ volume. }
\label{rings-camera}
\end{figure}

 Droplets dried on glass show a pattern which depends on the method of preparation of the sample. The dried droplets were photographed using Nikon CoolPix L120. Fig (\ref{rings-camera}) shows typical patterns generated. When the aqueous gelatin solution is prepared and dry powdered salt mixed with it, the patterns are as shown in Fig (\ref{rings-camera})(a) for salt concentration 0.01 M. A series of concentric rings are formed, surrounded by a transparent annular region. When the salt concentration is increased to 0.03 M, the pattern looks more or less similar but the transparent border is much reduced in width (fig \ref{rings-camera} (b)). 
 
 However, when the sample is prepared with the same constituents, but following a slightly different procedure, a new feature is observed in the pattern. Now the gelatin-water and sodium sulphate-water solutions are prepared separately and then mixed. In this case a radial fingering pattern appears, which is superposed on the rings (fig (\ref{rings-camera})(c) and (d)).  The fingering pattern is seen for low salt concentrations $\sim 0.02 M$ and it disappears when salt concentration is increased to $\sim 0.04 M$ or higher.

 \subsubsection{Medium resolution imaging}
\begin{figure}[h]
\begin{center}
\includegraphics[width=10.0cm, angle=0]{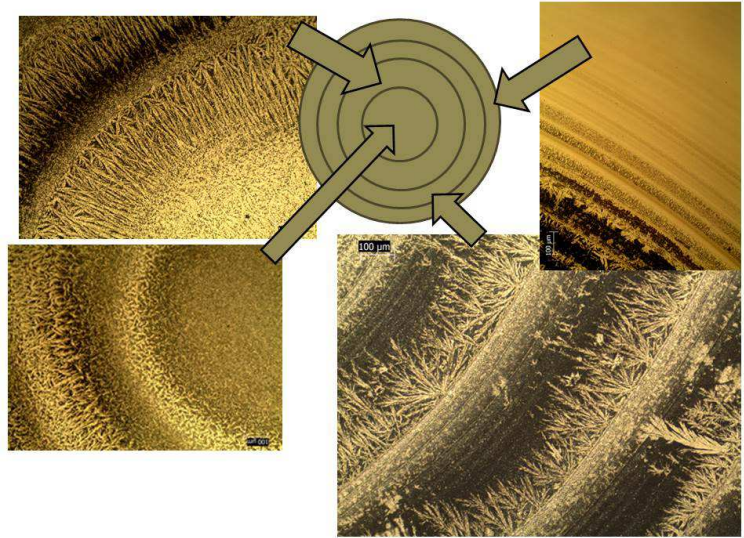}
\end{center}
\caption{Images of different regions of the pattern under microscope are shown. The scale bars in the figure panels are all 100 $\mu m$. Hierarchical ring patterns with dendritic crystal growth in between are observed.}
\label{micro-rings}
\end{figure}

 The rings shown in fig(\ref{rings-camera}) are actually \textit{bands} of several closely spaced thinner rings. The rings are not equispaced, but the spacing has a certain order. This fine structure is seen at higher resolution. Complex dendritic structures growing from the rings are also visible, when the drops are viewed under a microscope (Leica DM 750). Fig (\ref{micro-rings}) shows the details of the pattern, for different regions of the droplet.  What appeared to be a single rings in fig (\ref{rings-camera}) is seen to be sets of rings with dendritic crystals growing in between them. The fine structure of the different regions is shown in Fig (\ref{micro-rings}) where relative positions of the different micro-graphs are explained with respect to a schematic diagram of the whole pattern.
 
 \subsubsection{Scanning Electron Microscopy (SEM)}
\begin{figure}[h]
\begin{center}
\includegraphics[width=10.0cm, angle=0]{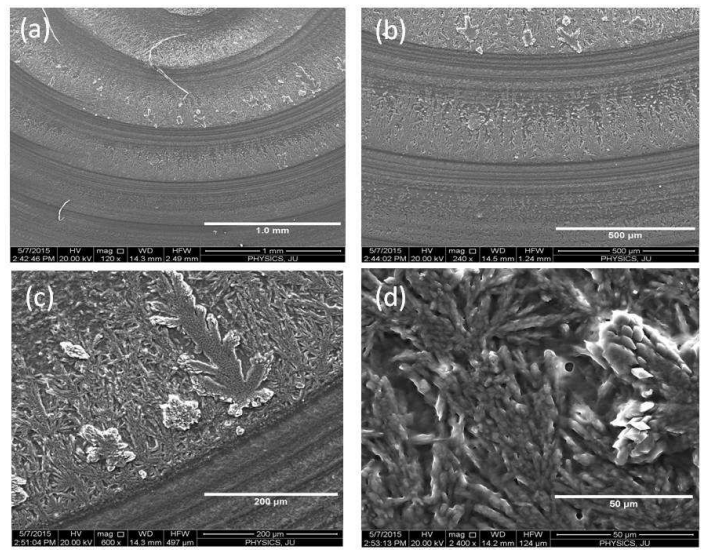}
\end{center}
\caption{SEM images taken at successively increasing resolution from (a) to (d) are shown. High resolution images of the dendrites show sodium sulphate crystallites.}
\label{semfig}
\end{figure}

 FESEM imaging has been done using SEM facility, configuration no. QUO-35357-0614 (funded by FIST-2,DST Government of India, at the Physics Department, Jadavpur University) to see the structure of the dendrites down to nano scales. The structures for increasing resolution are shown in fig (\ref{semfig}).
 Diamond shaped crystallites of sodium sulphate can be discerned at highest resolution as shown in fig (\ref{semfig}(d)).
 
 \section{Analysis of patterns}
\begin{figure}[h]
\begin{center}
\includegraphics[width=10.0cm, angle=0]{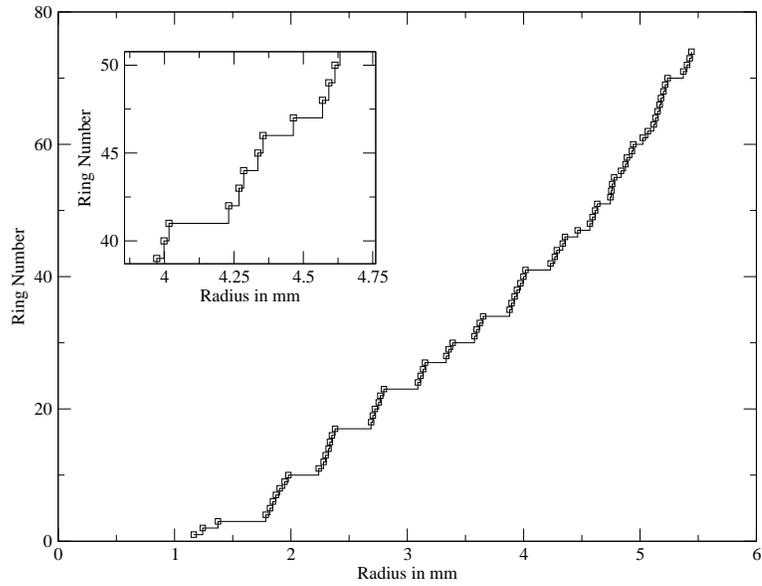}
\end{center}
\caption{The measured radii of the successively increasing rings are shown as a `staircase' pattern. The inset shows a zoomed up portion, demonstrating the self-similarity.}
\label{radius-exp}
\end{figure}

 The ring patterns have a typical hierarchical structure which looks self-similar for a certain range of length-scales. To analyse the pattern, the following procedure is adopted. The radial position of each ring with reference to the centre of the pattern is tabulated. Let us number the rings consecutively from the central reference point and denote the radius of the $n$th ring by $R_n$. We plot $n$ on the Y-axis against $R_n$ on the X-axis, as shown in fig (\ref{radius-exp}). If the points are joined to the Y-axis by horizontal lines, such that all points where $R_n<R_{n+1}$ are on the same `step' corresponding to $Y=n$, we get a staircase-like structure. It appears that as we observe the pattern at higher and higher resolutions new rings appear, that is new steps appear between the existing ones. So this pattern resembles a Devil's staircase where ideally an infinite number of steps exist between any two steps. For real systems of course, there is a finite cut-off.
 
\begin{figure}[h]
\begin{center}
\includegraphics[width=10.0cm, angle=0]{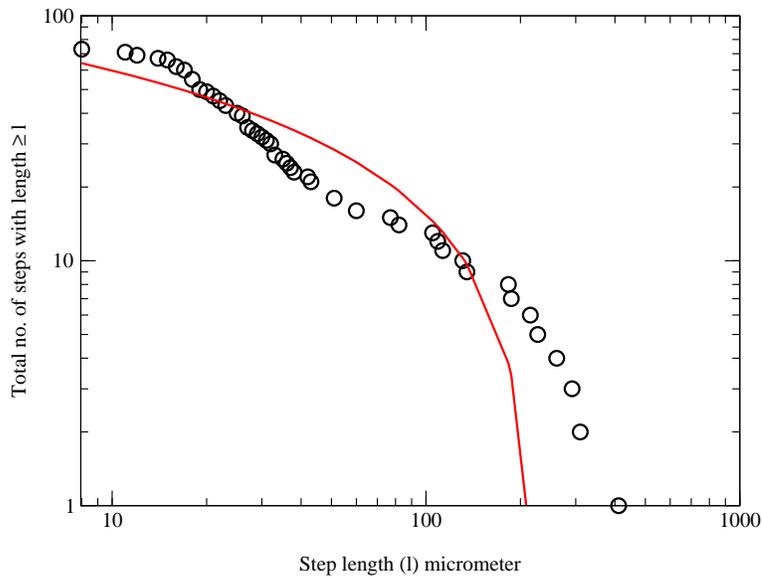}
\end{center}
\caption{The number of rings $N_l$ with length $\geq l$ are plotted against $l$. A logarithmic fit is shown in red.}
\label{expfit}
\end{figure}

 It is seen from fig (\ref{radius-exp}) that there are a few wide steps interspersed by many shorter steps. We plot a cumulative distribution of step lengths by counting the number $N_l$ of steps with length $\geq l$ and plotting $N_l$ against $l$. The resulting curve can be approximately fit by a logarithmic function of the form
 
 \begin{equation}
 Y = -A -B lnX
 \end{equation}
 as shown in fig (\ref{expfit}).
 \subsection{Hierarchical ring pattern: a simple model}
 
The hierarchical pattern of rings is a notable feature of the pattern and we offer a simple explanation for the process. Kaya et al.$^{5}$ have calculated the observed regular ring spacings for the system sodium poly(styrene
sulfonate) with sodium chloride. Their approach is as follows: the `coffee ring' at the outer periphery of the droplet is formed due to the highest evaporation rate at the outer boundary of the drop, which draws fluid carrying the solute towards the boundary. The solute particles get pinned at the boundary to form the so called `coffee-ring'. Kaya et al. suggest that the process may not stop here. The high concentration of solute at the boundary drives a diffusion flux inward, this superposes the outward advective flux and gives rise to non-monotonic concentration contour in the region  nearer the  droplet centre from the  coffee ring. Kaya et al.$^{5}$ calculate the shape of this concentration contour  and find a maximum, which exceeds the critical solute concentration for deposition. This leads to a second ring of smaller radius than the outer one and concentric with it. This process repeats to form a third ring, fourth ring and so on. In the model of Kaya et al. only the largest ring formed, participates in the diffusion process and the resulting rings are approximately equally spaced.

We offer a simple heuristic extension of the above scenario. We assume that when the fluid is complex and highly viscous, like gelatin, the outer rings do not dry out immediately and so they continue to participate in diffusion. We suppose that all, or at least several rings are active simultaneously. We show by a simplified argument that this leads to a system of rings whose spacings closely resemble the real pattern generated in our drying experiments.

\begin{figure}[h]
\begin{center}
\includegraphics[width=10.0cm, angle=0]{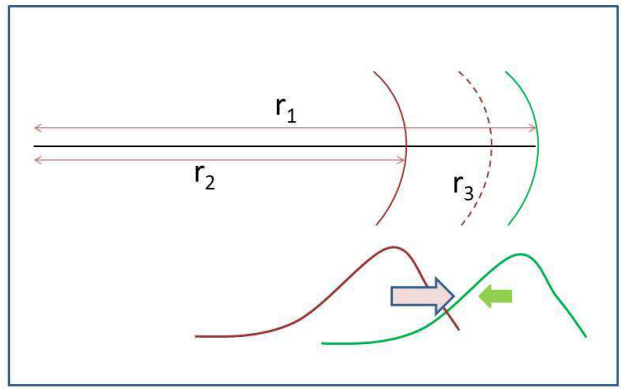}
\end{center}
\caption{Schematic diagram showing successive ring formation. The ratio $r_2/r_1$ = b. The contours below are approximate outlines of the variation in salt concentration with distance from the center (after Kaya et al.$^{5}$) and the arrows represent the opposing diffusion fluxes. A ring forms when the maximum of the contour exceeds the critical salt concentrstion.}
\label{ring-formation}
\end{figure}

From the concentration contour calculated by Kaya et al.$^{5}$, we see that the contour is asymmetric with a higher gradient outward (towards the ring boundary) and a lower gradient inward. So we may assume faster diffusion outward and slower inward and as a result when the two concentration fluxes superpose, the maximum of the resultant contour will be shifted towards the larger ring, instead of forming exactly in the middle. Let us start with an outer ring at radius $r_1$ and a subsequent one at $r_2$ (see fig \ref{ring-formation}), where the radii are related by a constant ratio $b = r_2/r_1$. We assume the next ring to form at $r_3$, due to the superposing diffusion fluxes from rings 1 and 2, maintaining the same ratio.
So

\begin{equation}
b=\frac{r_2}{r_1}= \frac{r_3-r_2}{r_1-r_2}
\end{equation}

this process follows for successive rings which form closer and closer to $r_1$, for the $n$th and $(n+1)$th such rings we have then

\begin{equation}
b = \frac{r_{n+1}-r_n}{r_1-r_n}
\end{equation}

This allows us to write a recurrence relation for the radius of successive rings as

\begin{equation}
r_{n+1} = r_n(1-b)+ br_1
\label{recc}\end{equation}
for $n>1$.

 We can thus generate a sequence from two initial parameters $b$ and $r_1$
 Similar such sequences can be written taking the initial ring radius as $r_2$, with $n>2$ and so on  with the ratio $b$ replaced by other constants say $c$, $d$ etc.
 
  Then the next sequence is given by
 \begin{equation}
r_{m+1} = r_m(1-c)+ cr_2 
\label{rec2}
\end{equation}
for $m>2$.

\begin{figure}[h]
\begin{center}
\includegraphics[width=10.0cm, angle=0]{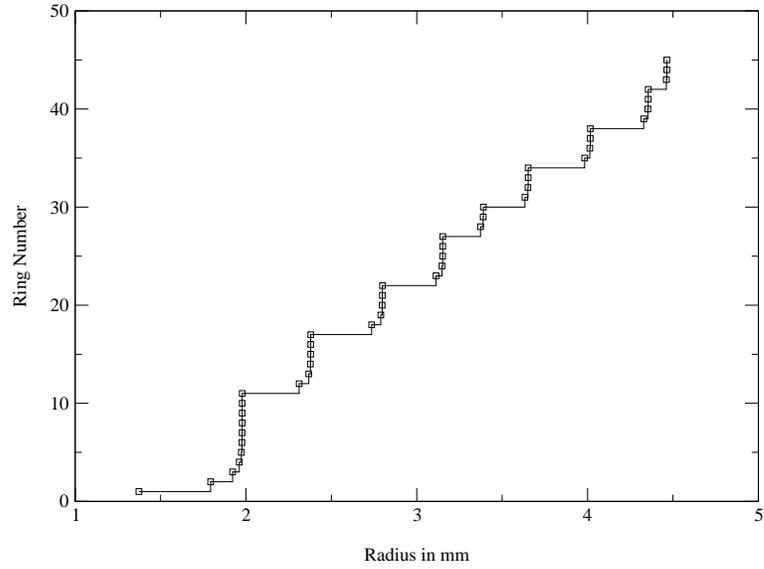}
\end{center}
\caption{Radii of rings calculated from eqs (\ref{recc} and \ref{rec2}) are shown, here the ratio $b$ is successively varied between 0.6 to 0.9.}
\label{varb-rings-calc}
\end{figure}

\begin{figure}[h]
\begin{center}
\includegraphics[width=10.0cm, angle=0]{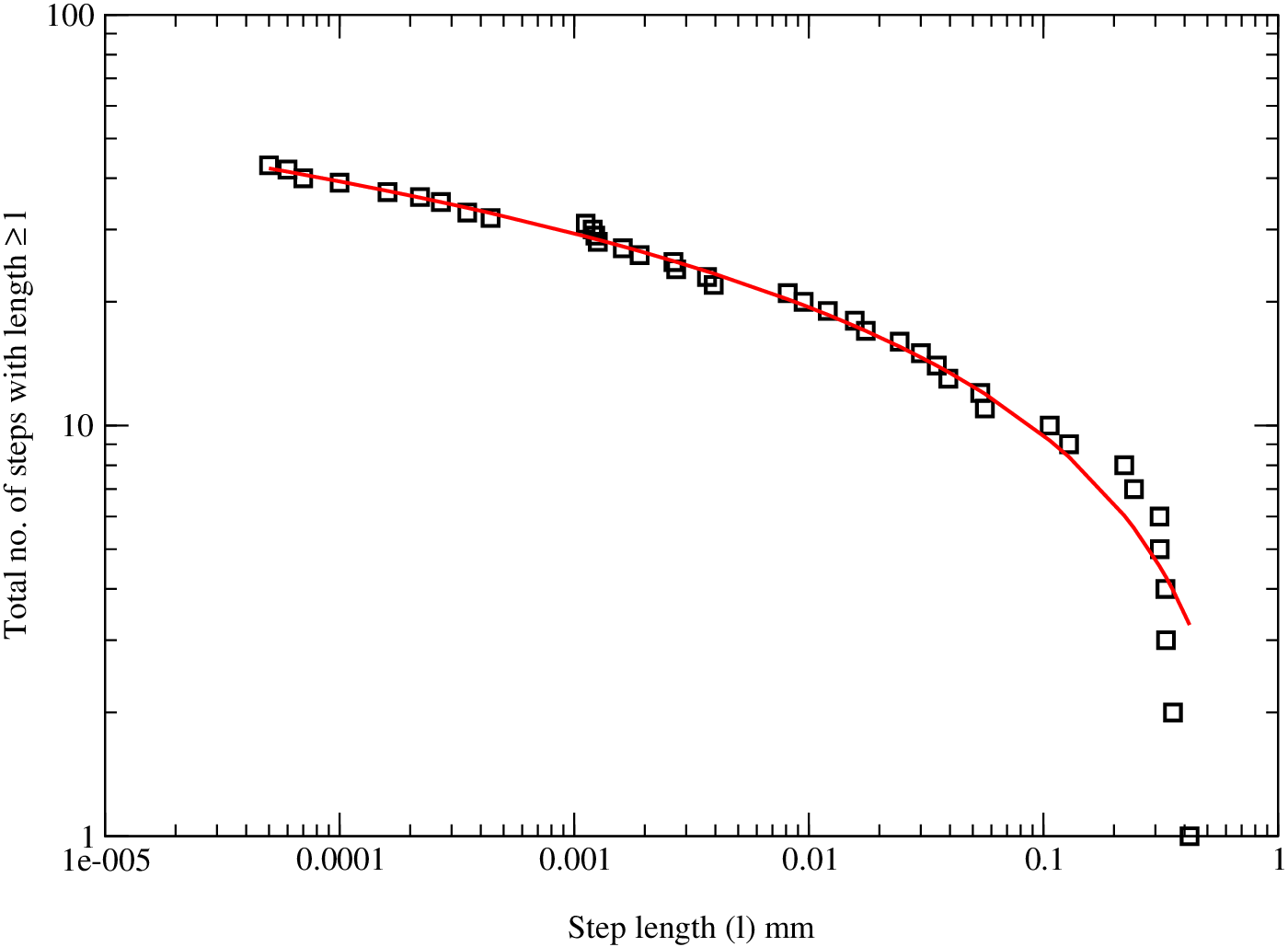}
\end{center}
\caption{The number of rings $N_l$ with length $\geq l$ from fig(\ref{varb-rings-calc}) are plotted against $l$. A logarithmic fit is shown in red. }
\label{varb-fit}
\end{figure}

Plotting the positions of several such series we get a ring pattern like fig (\ref{varb-rings-calc}), where $b$ is varied successively from 0.6 to 0.9 (which is close to the experimental results). It can be seen that the resulting pattern of rings qualitatively reproduces the experimental patterns (fig \ref{radius-exp}). In the real system, dendrites grow in the space between the adjacent rings, as shown with increasing magnification in SEM images (fig \ref{semfig}). We calculate the cumulative number of steps $N_l$ greater than a length $l$ and plot it against $l$, as done for the experimental ring sequence. The results can be fit approximately to a logarithmic plot as shown in fig (\ref{varb-fit}). So this very simple picture gives a qualitative description of the experimental patterns.

\subsection{Viscous fingering pattern}

The low resolution images shown in fig(\ref{rings-camera}) show fingering patterns superposed on the rings, when the solution is prepared by Method II. It is well known that viscous fingering patterns are formed when a low viscosity fluid, at a higher pressure is forced into a higher viscosity fluid, kept at a lower pressure$^{11}$. The two fluids should be immiscible.

The formation of the fingering pattern in the drying droplet may have the following origin. When Method II is employed the fluid (A) - a solution of gelatin in water is mixed with the solution (B) which is a solution of the salt in water. As the droplet dries, there is a tendency for the solutions to separate, this is seen by the formation of the transparent borders in fig(\ref{rings-camera}), which are mostly gelatin. Viscosity measurement shows that, before drying starts  the salt solution has a lower viscosity ($\sim 1cp$) compared to the gelatin solution, which has viscosity ($\sim 1.6 cp$). As drying starts, a pressure gradient develops with a lower pressure at the drop periphery, due to higher evaporation rate here. So the condition for viscous fingering is satisfied and patterns form. Surface tension measurements show that the two fluids have different surface tensions and are therefore immiscible. Though the fluids appear to form a homogeneous solution when mixed together, on a microscopic level they retain their identity as in an emulsion.

At higher salt concentrations, salt becomes more uniformly distributed, there is no gelatin border and the fingering pattern disappears. Following the preparation Method I also shows no fingering. In this case two different fluids are not mixed, as the gelatin is added in powder form to the salt solution, so viscous fingering does not occur.

\begin{table}
\begin{center}
\begin{tabular}{|c|c|c|}
\hline
Sample  & Surface Tension   &     Viscosity \\& (mN/m) &(cp) \\ \hline   
Gelatin solution & 39.0 & 2.06 \\
Na$_2$SO$_4$ solution & 70.0 & 1.65 \\Mixture of Gelatin and Na$_2$SO$_4$ solution & 40.6& 2.15\\
\hline

\end{tabular}
\caption{Measured physical properties of the experimental solutions at 30$^\circ$C and relative humidity 52\%} \label{prop}
\end{center}
\end{table}

 \section{Discussion}
 The study of drying drops has become a highly popular topic in recent times$^{12,7,6,13}$.  The beauty of the fascinating patterns formed is not the only reason for this widespread interest. Medical diagnostics$^{12}$ promises to be an area where patterns of dried biological fluids will be of practical use. Other practical applications are in ink-jet printing and drying of thin coatings.
 
 We discuss here two aspects of these patterns which to our knowledge, have not yet been reported or explained. Though concentric rings have been reported in drying droplets, the typical pattern we observe has not yet attracted much attention. The simultaneous presence of viscous fingering and ring-dendrite patterns appears to be a new observation as well. The models we propose here are very elementary and need to be developed for a complete understanding of these phenomena. We hope to pursue these studies and report work on more detailed models in future.

\section{Acknowledgements}  Authors are grateful to DST, Govt. of India for funding this research through project No.SR/S2/CMP-127/2012. B. Roy thanks DST for providing a Junior Research fellowship and M. Dutta Choudhuri thanks CSIR for granting a Senior Research Fellowship. DST, Govt. of India is acknowledged for funding the FESEM facility at Physics Department, Jadavpur University.

\section{References}

[1]C.C. Annarelli, J. Fornazero, J. Bert, and J. Colombani, \textit{ Eur. Phys. J. E},2001, 5, 599-603\\
 \noindent
[2] P. Takhistov, H-C Chang, \textit{Ind. Eng. Chem. Res.}, 41, 2002, 6256-6269.\\
 \noindent
[3] Moutushi Dutta Choudhury, Tapati Dutta, Sujata Tarafdar, \textit{Colloids and Surfaces A:Physico chem. Eng. Aspects}, 2013, 432, 110–118\\
\noindent[4] Robert D. Deegan, Olgica Bakajin, Todd F. Dupont, Greg Huber, Sidney R. Nagel, and Thomas A. Witten, \textit{Physical Review E}, 2000, 62, 756-765.\\
\noindent[5] D. Kaya, V. A. Belyi, and M. Muthukumar, \textit{The Journal of Chemical Physics}, 2010, 133, 114905, doi: 10.1063/1.3493687\\
\noindent[6] Y.Y. Tarasevich, \textit{Phys. Rev. E}, 2005, 71, 027301\\
\noindent[7] YongJian Zhang,  ZhengTang Liu, DuYang Zang, YiMeng Qian, KeJun Lin, \textit{SCIENCE CHINA Physics, Mechanics \& Astronomy}, 2013, 56, 1712–1718\\
\noindent[8] Carlos Rodriguez-Navarroa, Eric Doehne, Eduardo Sebastian,
\textit{Cement and Concrete Research}, 2000, 30, 1527 - 1534\\
\noindent[9]  Andrea Hamilton, Christopher Hall, Leo Pel, \textit{J. Phys. D: Appl. Phys}, 41 (2008) 212002 (5pp)\\
\noindent[10] B.B. Mandelbrot, \textit{Fractal Geometry of Nature, 
}, 1977, W.H. Freeman \& Co., New York\\
\noindent[11] T. Vicsek, \textit{Fractal Growth Phenomena}, 2nd ed.,1992, World Scientific, Singapore\\
\noindent[12] \textit{Droplet Wetting and Evaporation:From Pure to Complex Fluids}, ed. D. Brutin, 2015, Academic press, ISBN: 978-0-12-800722-8\\
\noindent [13] N.S. Bonn, S. Rafai, D. Bonn, G. Wegdam, \textit{Langmuir}, 2008, 24, 8599-8605\\

\end{document}